# Preoperative Rotator Cuff Tear Prediction from Shoulder Radiographs using a Convolutional Block Attention Module-Integrated Neural Network

Authors:

Chris Hyunchul Jo[1†], Jiwoong Yang[2†], Byunghwan Jeon[3], Hackjoon Shim[4,5], Ikbeom Jang[3]*

Affiliations:

[1] Department of Orthopedic Surgery, Seoul National University College of Medicine, SMG-SNU Boramae Medical Center, Seoul, South Korea

[2] Artificial Intelligence Semiconductor, Hanyang University, Seoul, South Korea

[3] Division of Computer Engineering, Hankuk University of Foreign Studies, Yongin, South Korea

[4] Medical Imaging AI Research Center, Canon Medical Systems Korea, Seoul, South Korea

[5] ConnectAI Research Center, Yonsei University College of Medicine, Seoul, South Korea

Presenting author:

Full name: Chris Hyunchul Jo

Affiliation: Department of Orthopedic Surgery, Seoul National University College of Medicine, SMG-SNU Boramae Medical Center, Seoul, South Korea

Email: chrisjo@snu.ac.kr

Key information:

Research question: We test whether a plane shoulder radiograph can be used together with deep learning methods to identify patients with rotator cuff tears as opposed to using an MRI in standard of care.

Findings: By integrating convolutional block attention modules into a deep neural network, our model demonstrates high accuracy in detecting patients with rotator cuff tears, achieving an average AUC of 0.889 and an accuracy of 0.831.

Meaning: This study validates the efficacy of our deep learning model to accurately detect rotation cuff tears from radiographs, offering a viable pre-assessment or alternative to more expensive imaging techniques such as MRI.

†: equal contribution, *: corresponding author

MANUSCRIPT

Introduction

Initial radiograph evaluations often fail to identify soft tissue injuries such as rotator cuff tears. It necessitates further imaging with more expensive MRI examinations, increasing healthcare costs. In this study, we show that a convolutional neural network with channel attention and spatial attention modules can significantly enhance the accuracy of rotator cuff tear detection, only using a single shoulder radiograph. All shoulder radiograph data used for training the deep learning model were collected from our local clinic. It may offer a viable pre-assessment or alternative to more expensive imaging techniques such as MRI.

Material and methods

**Data**

We retrospectively collected a dataset of shoulder radiographs from 99 patients from our clinic. The dataset comprises 50 patients with full-thickness rotator cuff tears (fRCT) and 49 without tears. We acquired radiographs in four angles – axial, glenoid, outlet, and anteroposterior (AP) – totaling 396 images. All the images were acquired before surgery. Regions of interest (ROIs) essential for fRCT diagnosis were identified and annotated with bounding boxes on all images. These annotations were used to train the YOLO v5 model to automatically crop ROIs from all radiographs, and to efficiently handle the annotation process for all future data. The results fo ROI extraction are shown in Figure 2 and Table 1. The dataset was divided based on subject IDs with each containing four view-specific images. We applied 5-fold cross-validation due to the limited size of the dataset, resulting in 316 training images from 79 subjects and 80 test images from 20 subjects, ensuring no subject overlap between folds.

**Network Architecture and Training**

The ROIs extracted were further processed using Contrast-Limited Adaptive Histogram Equalization (CLAHE) to enhance bone structures and edge visibility, facilitating more detailed recognition of fractures and structural integrity. We employed a ResNet50 model applied with a Convolutional Block Attention Module (CBAM) to diagnose rotator cuff tears. CBAM enhances the model's learning focus on essential features in medical image by sequentially applying channel attention and spatial attention. This method allows for a concentrated analysis of salient features crucial for accurately diagnosing rotator cuff tears. The pretrained ResNet50 was adapted to classify between the presence and absence of fRCT, with only two output classes in the final layer. The architecture of the model is shown in Figure 4. Due to the limited dataset of 99 subjects, we employ k-fold cross-validation to enhance the accuracy of the model's performance. Each fold in the k-fold cross-validation setup was balanced to have an equal ratio of fRCT and no-tear cases in both the train and validation datasets.

**Implementation Details**

We trained models on an NVIDIA RTX 3090 GPU using the SGD optimizer with a learning rate of 0.01 and a batch size of 8. To prevent overfitting, we employed various data augmentation techniques including rotation, horizontal flipping, random crop, scaling, translation, brightness adjustment, and inversion, as well as implementing a dropout rate of 0.2. All data was resized to 512x512 pixels prior to training. We utilized a CrossEntropyLoss for the loss function and a CosineAnnealingWarmupRestarts scheduler to dynamically adjust the learning rate during the training process.

Results

Using k-fold cross-validation, we evaluated the performance of our model across each fold. The average accuracy achieved was 0.831 (329/396). The AUROC for the two classes was 0.889, indicating a high level of discriminative ability. The Positive Predictive Value (PPV) was 0.852 (161/189), and the Negative Predictive Value (NPV) was 0.812 (168 /207). This is shown in Figure 5. The findings demonstrate that radiographs alone can effectively classify patients with fRCT, underscoring its potential utility in diagnostic settings.

Discussion and Conclusion

Our proposed method demonstrates that radiographs alone can effectively diagnose fRCT. This approach calculates the probability of rotator cuff tears, thereby assisting in the decision-making process of whether to proceed with MRI imaging. To increase the generalizability of the model, we plan to collaborate with multiple centers to incorporate a larger and more diverse set of radiograph data. Future research will aim to expand the dataset, which was collected from our clinic, and analyze the impact of different radiographic views (axial, glenoid, outlet, and AP) on the diagnosis of rotator cuff tears to improve the current model's performance. In the long term, we aim to achieve a level of diagnostic accuracy close to that of MRI-based diagnostics for rotator cuff tears.

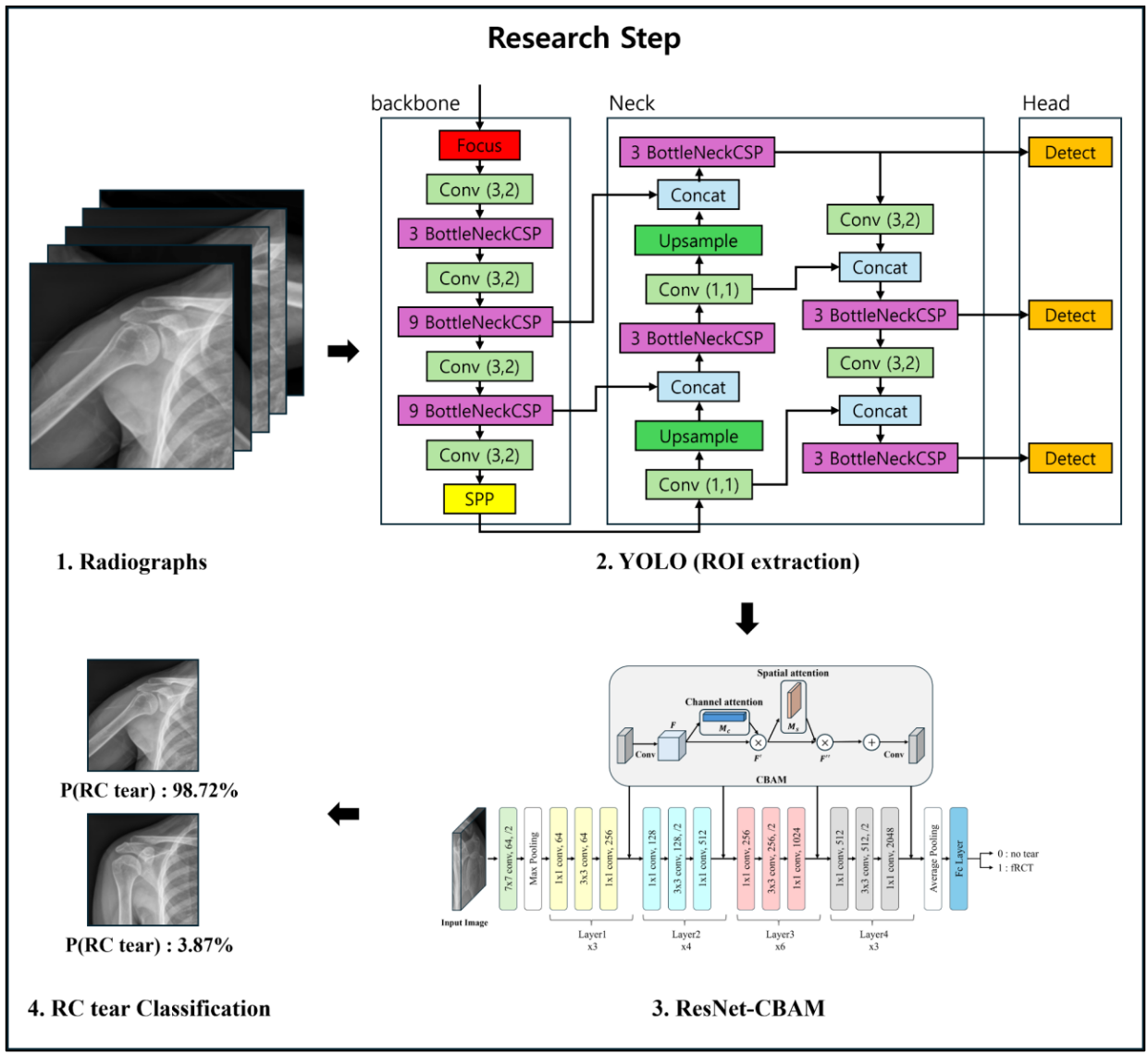

**Figure 1**: Stages of a study for diagnosing rotator cuff tears using radiographs. The radiographs are processed through Yolo v5 to extract essential Regions of Interest (ROI). These are then used to train the ResNet-CBAM model, enabling the classification of patients with rotator cuff tears. Our study proposes a diagnostic framework that predicts the presence of rotator cuff tears using only radiographs, a more cost-effective and accessible option. The proposed method calculates the probability of rotator cuff tears in patients, thereby assisting in the decision-making process of whether to proceed with MRI imaging.

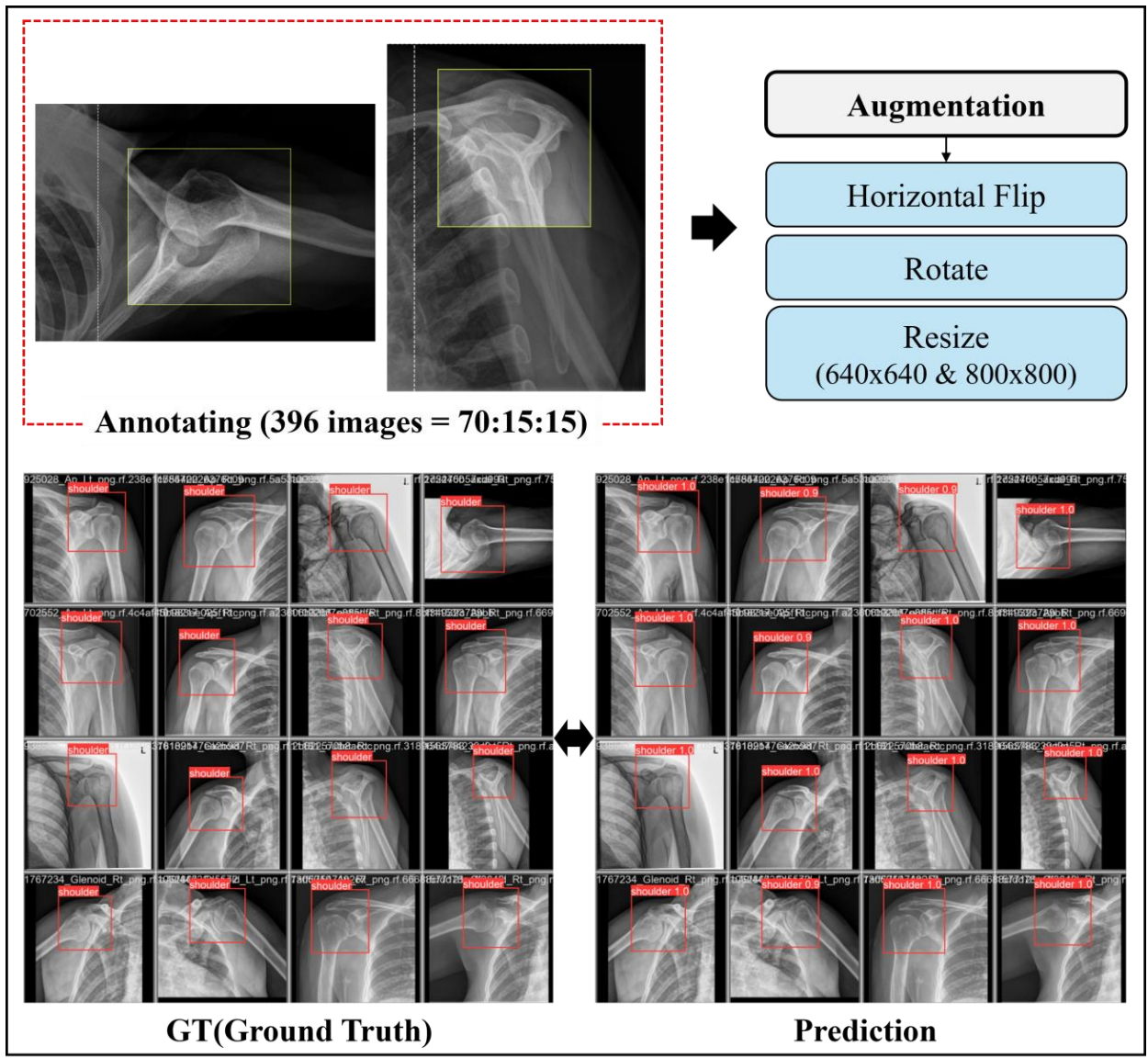

**Figure 2**: Training YOLO v5 for ROI Extraction. We conducted the process of annotating 396 radiograph images from 99 subjects, followed by training the YOLO v5 model to detect regions of interest (ROI). The data was divided into training, validation, and test sets in a 70:15:15 ratio. The annotated data was subjected to augmentation processes including horizontal flip, rotation, and resizing. For resizing, the original aspect ratio of the images was maintained, and the remaining space was filled with black edges. Two versions of the model were trained: version 1 (exp1) with images resized to 640x640 pixels and version 2 (exp2) with images resized to 800x800 pixels. The best-performing version, exp 2, was selected based on its superior performance in ROI extraction as indicated in Table 1. The ground truth (GT) and prediction results illustrate that the YOLO v5 model successfully detected th ROIs with high accuracy, demonstrating its effectiveness in this application.

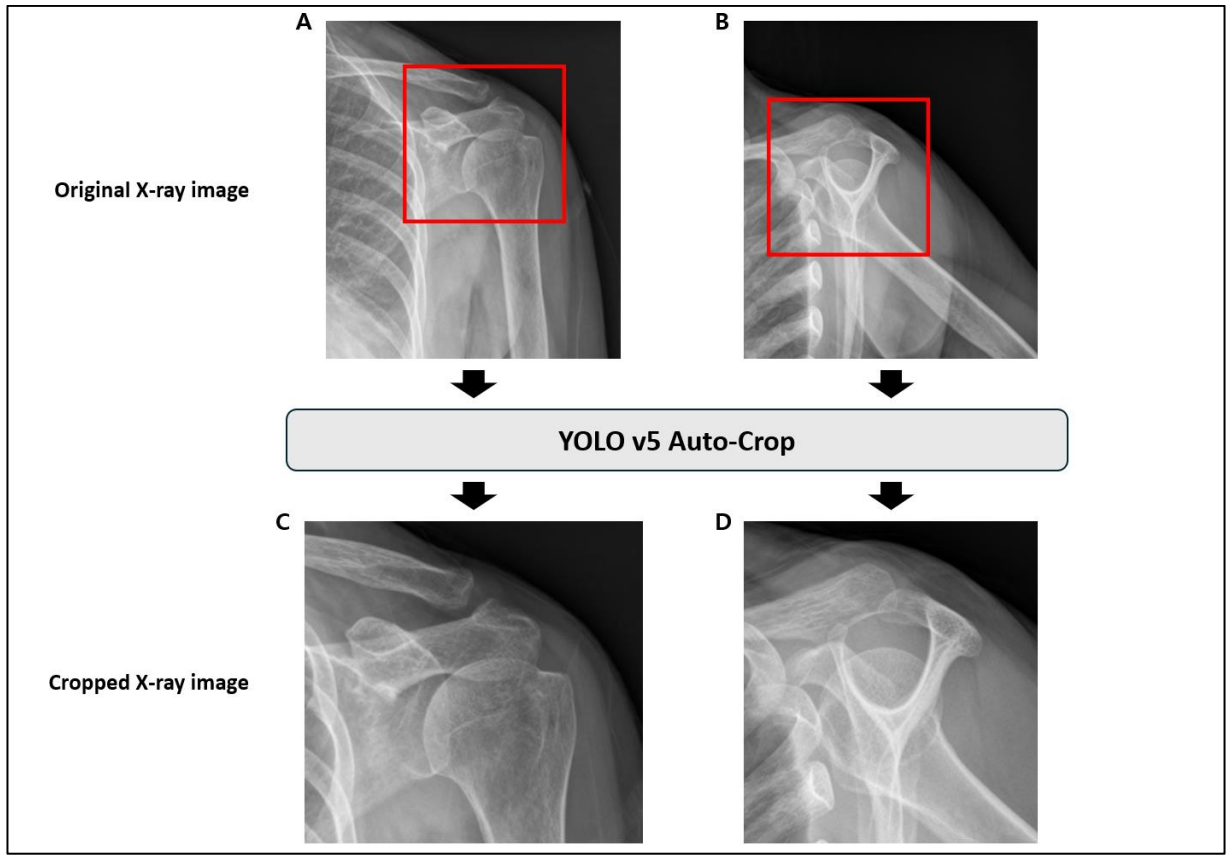

**Figure 3**: Automated ROI Cropping in radiographs Using YOLO v5. We train YOLO v5 with data labeled with bounding boxes for the ROI. When the saved best-weights (exp2) are loaded and applied, all radiographic data in the dataset are automatically cropped to include only the ROI. For the diagnosis of fRCT, specific regions of interest (ROIs) were cropped from the original-sized radiographs A and C to produce images B and D, respectively. These particular ROIs were then utilized as training data. This process allows the model to more effectively identify the features of full-thickness Rotator Cuff Tears (fRCT) by excluding unnecessary regions during training. Additionally, it reduces variations in image size and positioning that can occur depending on the individual who performed the radiograph, thereby aiding in generalization.

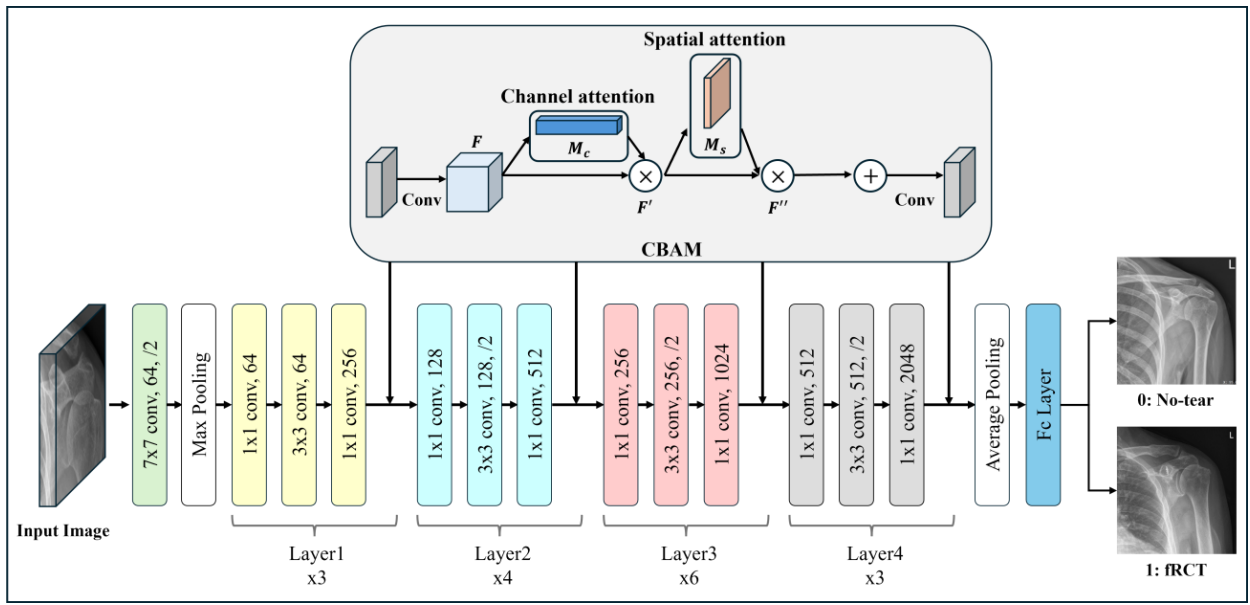

**Figure 4**: ResNet50-CBAM model structure. This figure is ResNet50 architecture with CBAM applied, enhancing feature representation by applying sequential channel and spatial attention mechanisms. It aids in enhancing performance by focusing on features necessary for the diagnosis of fRCT.

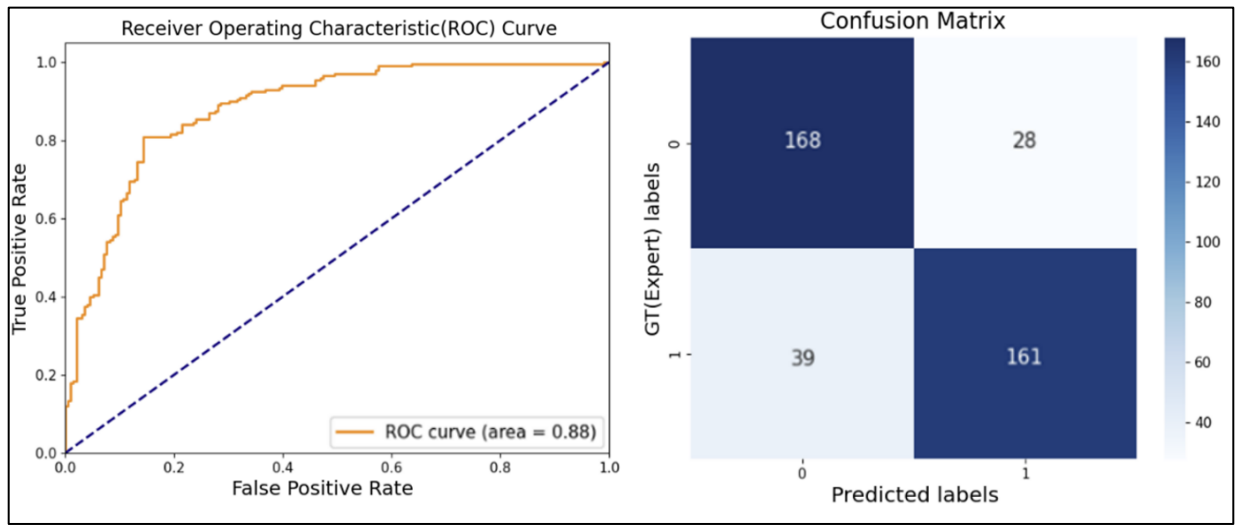

**Figure 5**: ROC curves(A) and confusion matrix(B) on the test set. In the confusion matrix, 0 represents 'no-tear' and 1 indicates 'fRCT'. During k-fold cross-validation, the cumulative results predicted by the model in each fold's test phase are presented in a confusion matrix. This is useful for assessing the average performance metrics of the dataset.

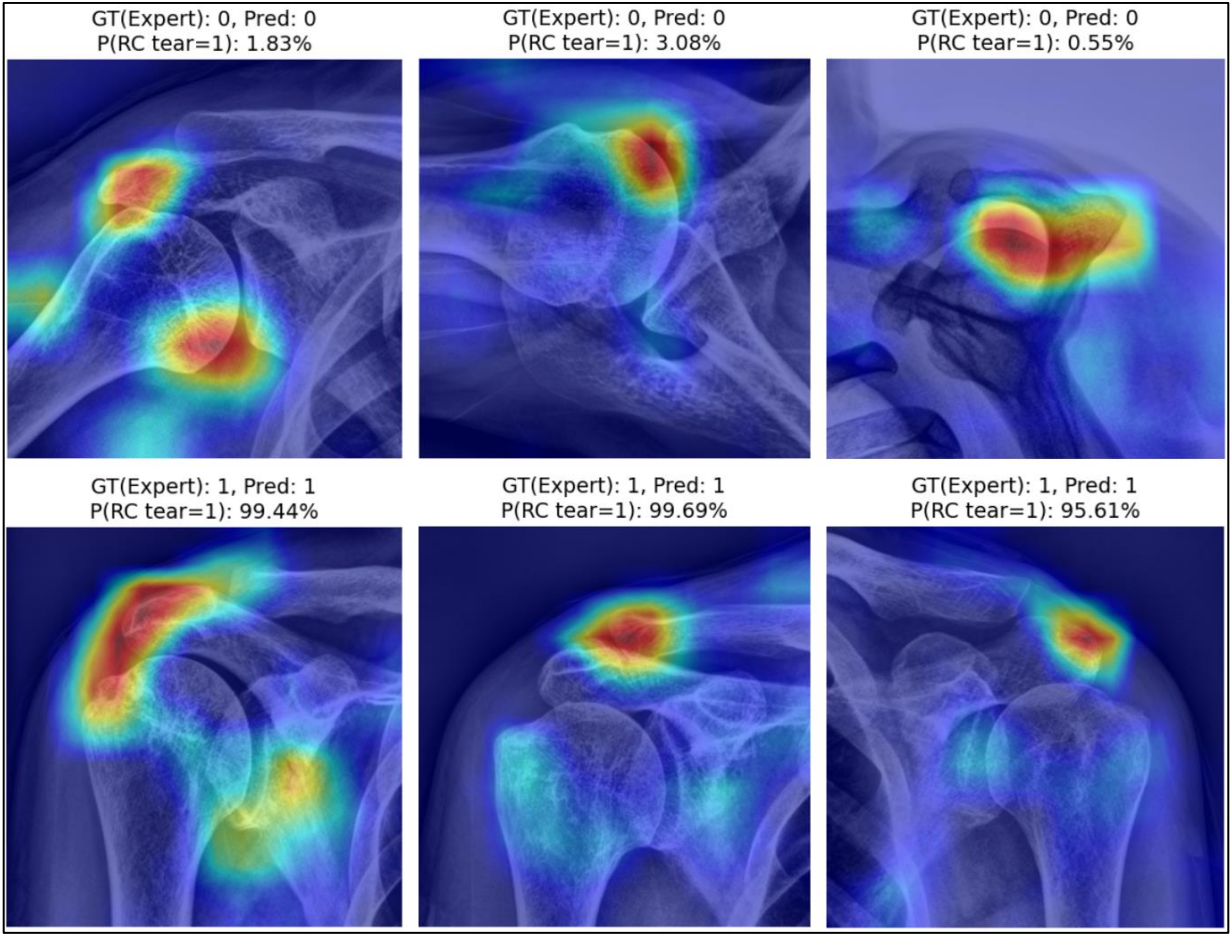

**Figure 6**: Visualization of Prediction Results Using Grad-CAM. We visualized which regions of the radiograph the model focused on during training using Grad-CAM. It was observed that the areas our model concentrated on through learning closely align with the regions that should be examined during the initial diagnosis stages of fRCT.

**Table 1: Comparison of YOLO v5 Training Experiments.**

The metrics used to evaluate the YOLO v5 model include Precision, Recall, mAP_0.5, and mAP_0.5:0.95. mAP_0.5 measures the mean Average Precision at an Intersection over Union (IoU) threshold of 0.5, while mAP_0.5:0.95 averages the mean Average Precision across multiple IoU thresholds ranging from 0.5 to 0.95. In our experiments, exp1 resized images to 640x640 pixels, and exp2 resized images to 800x800 pixels. As shown in Table 1, exp2 slightly outperformed exp1 in mAP_0.5:0.95, achieving 0.8003 compared to exp1's 0.7984. Therefore, we used the best weights from exp2 to perform the ROI crop on the images.

| Experiment version | Precision | Recall | mAP_0.5 | mAP_0.5:0.95 |
|---|---|---|---|---|
| exp1 | 0.9975 | 1.000 | 0.9950 | 0.7984 |
| **exp2** | **0.9986** | **1.000** | **0.9950** | **0.8003** |